\documentclass[prl,amsmath,amssymb,aps,showkeys,nofootinbib,twocolumn]{revtex4-2}

\usepackage{aas_macros} 
\usepackage{graphicx}
\usepackage[dvipsnames]{xcolor}
\usepackage{hyperref}
\usepackage{microtype}
\usepackage{dcolumn}

\newcommand{\Mh}{\ensuremath{h^{-1}M_{\odot}}}

\newcommand{\Mpch}{\ensuremath{h^{-1}{\rm Mpc}}}

\newcommand{\be}{\begin{equation}}
\newcommand{\ee}{\end{equation}}

\usepackage{multirow}
\usepackage{tabularx}
\newcommand{\skeletor}{\texttt{Skeletor}}
\usepackage[normalem]{ulem}

\begin{document}

\title{Filaments within filaments: unraveling the hierarchy of the cosmic web}

\author{Saee Dhawalikar}
\altaffiliation{Corresponding author}
\email{saee@usal.es}
\affiliation{Departamento de Física Fundamental, Universidad de Salamanca, E-37008 Salamanca, Spain}
\affiliation{Inter-University Centre for Astronomy \& Astrophysics, Ganeshkhind, Post Bag 4, Pune 411007, India}

\author{Aseem Paranjape}
\affiliation{Inter-University Centre for Astronomy \& Astrophysics, Ganeshkhind, Post Bag 4, Pune 411007, India}

\author{Shadab Alam}
\affiliation{Tata Institute of Fundamental Research, Homi Bhabha Road, Mumbai 400005, India}

\date{\today}

\begin{abstract}
Cosmic filaments are the most striking feature of the hierarchical cosmic web. Yet, their own sub-structure remains unexplored, unlike the well-known, universal halo/sub-halo hierarchy. We use a hierarchical reconstruction scheme to identify sub-filaments within larger filaments. Examining differences in the radial phase-space profiles of parent and sub-filaments, we argue that the latter constitute a statistically distinct class of structures. Establishing the existence of a universal filament hierarchy must then account for these differences, with implications for cosmology and dark matter. 
\end{abstract}


\maketitle

\noindent
\textbf{\emph{I. \underline{Introduction:}}}\\
A key prediction of the $\Lambda$-cold dark matter ($\Lambda$CDM) paradigm is the near-universal, hierarchically organized phase-space structure of virialized halos \citep{NFW97}. 
Hierarchical growth through accretion and mergers produces not only isolated halos but also sub-halos embedded within larger host halos. Treating sub-halos as a distinct population has proven essential for understanding their structural evolution, tidal stripping, merger histories, and the connection between dark matter halos and galaxy populations \citep{Springel+2001, Cooray&Sheth2002,Wechsler&Tinker2018}, leading to a more complete description of hierarchical structure formation. 

One naturally expects a similar universal hierarchy to also manifest in the most prominent components of the cosmic web, \emph{viz.} cosmic filaments. The search for this hierarchy, however, has been stymied by the lack of well-established physical and numerical definitions of hierarchical structure in these objects. Instead, filament studies have typically treated all filaments as belonging to a single population \citep[e.g.,][]{Espinosa+2022,Xu+2026}.
Filament-like features embedded in the gravitational potential of a larger `parent' filament plausibly experience different tidal fields and matter flows than an isolated filament, ignoring which might generate avoidable scatter in filament statistics and hamper the search for universality.

Recently, we introduced \skeletor, a hierarchical filament finder that reconstructs the multi-scale cosmic web using point-wise tracer information and explicitly identifies a nested sequence of filamentary structures (see Dhawalikar \emph{et al.}, in preparation, 2026a, henceforth paper I and \citep{Saee_thesis}). The embedded structures in this hierarchy, which we refer to as \emph{sub-filaments}, emerge naturally at lower tracer masses, as increasingly smaller scales of the cosmic web are resolved. 

In this \emph{Letter}, we depart from the  standard practice of clubbing all filaments together and separately study sub-filaments and their parents statistically. 
Rather than a comprehensive characterization of sub-filament properties, our goal here is to perform an initial exploration of the idea that treating parent and sub-filaments as separate populations could lead to a cleaner and more physically motivated statistical description of the filamentary network. To this end, we compare their node properties and phase-space profiles. For simplicity, we adopt a sub-filament definition based on geometric overlap between filament spines and the radial extent of larger filaments. We note, however, that other definitions incorporating dynamical or evolutionary information may ultimately provide a more physically motivated classification, which we leave to future work.

Below, we start with the technical details of identifying filament samples in cosmological $N$-body simulations, followed by details of the parent and sub-filament classification leading up to the main discussion of sub-filament properties. We close with a discussion of future prospects.

\vskip 0.1in \noindent
\textbf{\emph{II. \underline{Filament samples:}}}\\
We use five independent realizations of $N$-body $\Lambda$CDM simulations with identical cosmological parameters, each evolving $1024^3$ particles in a periodic box of side length $300\Mpch$ with a particle mass of $\sim2\times10^9\Mh$. 
The simulations were evolved using \textsc{Gadget-2} \citep{GADGET2005}, halos were identified using the phase-space halo finder \textsc{Rockstar} \citep{Rockstar2013}, and the halo catalogues were subsequently cleaned as described in paper I.
For further details of the simulations and data products, kindly refer to Ref.~\cite{phs18}. We work at redshift $z=0$ throughout.
To improve statistics, the (sub-)filament samples from all independent realizations are pooled together for measuring profiles and distributions.

We identify filaments using \skeletor, which exploits the spatial anisotropy of Voronoi cells to identify cells preferentially compressed along one direction, characteristic of filamentary environments, and subsequently connects neighbouring filament-like cells into continuous structures. 

A key feature of \skeletor\ is that the reconstruction is performed hierarchically. Starting from a sparse population of the most massive tracers, the dominant filamentary network is first identified. The tracer mass threshold is then progressively lowered, revealing smaller-scale structures. We refer to the reconstruction at a given pair of node and tracer mass thresholds as a \emph{hierarchy level}. The primary level therefore contains the dominant, large-scale filaments, while subsequent levels recover progressively smaller structures. The initially reconstructed filament spine is first smoothed in Fourier space by applying a low-pass filter. It is subsequently refined using the surrounding dark matter distribution by globally minimizing the transverse mass asymmetry around the filament, producing a spine that more closely traces the center of the underlying matter distribution.

We run \skeletor\ to reconstruct three levels of filamentary hierarchy. As described in paper I, the portions of every filament lying within one halo radius  of the endpoint nodes are removed before any profile measurements are performed. We further discard filaments whose remaining length is smaller than the sum of the radii of the two endpoint nodes. As an additional cleaning step, we discard filaments whose enclosed density never exceeds $10\times$ the mean matter density.\footnote{As seen in paper I, filaments typically reach enclosed densities of several $10\times$ the mean matter density near their centers, with the outskirts of filaments typically reaching $\lesssim 10 \times$ the mean density. Filaments that fail this requirement, and which we remove, are unlikely to represent robust detections.} As a final cleaning step, we discard the $10\%$ of filament segments with the highest curvature, which can bias the measured radial profiles \citep[][paper I]{Dhawalikar&Paranjape2024}. The dependence of the phase-space structure on filament curvature is examined explicitly in Dhawalikar \emph{et al.}, (in preparation, 2026b, henceforth paper II) and \citep{Saee_thesis}. After these cuts, $82\%$ of the original catalog is retained, corresponding to 27789 filaments. These are then classified into parent and sub-filaments.

\vskip 0.1in \noindent
\textbf{\emph{III. \underline{Parent and sub-filament classification:}}}\\
For every filament identified at a given hierarchy level,
we compare its spine with the filament population reconstructed at all previous levels. For each current-level filament, we compute the fraction of its length lying within the estimated radius of previous-level filaments. The radius is taken to be the location of the radial velocity dip, the details of whose determination are described in paper II. If more than $50\%$ of the filament spine lies inside a previously identified filament, it is marked as a \emph{candidate sub-filament}.

\begin{figure}
\centering
\includegraphics[width=0.9\linewidth,trim=8 7 9 9,clip]{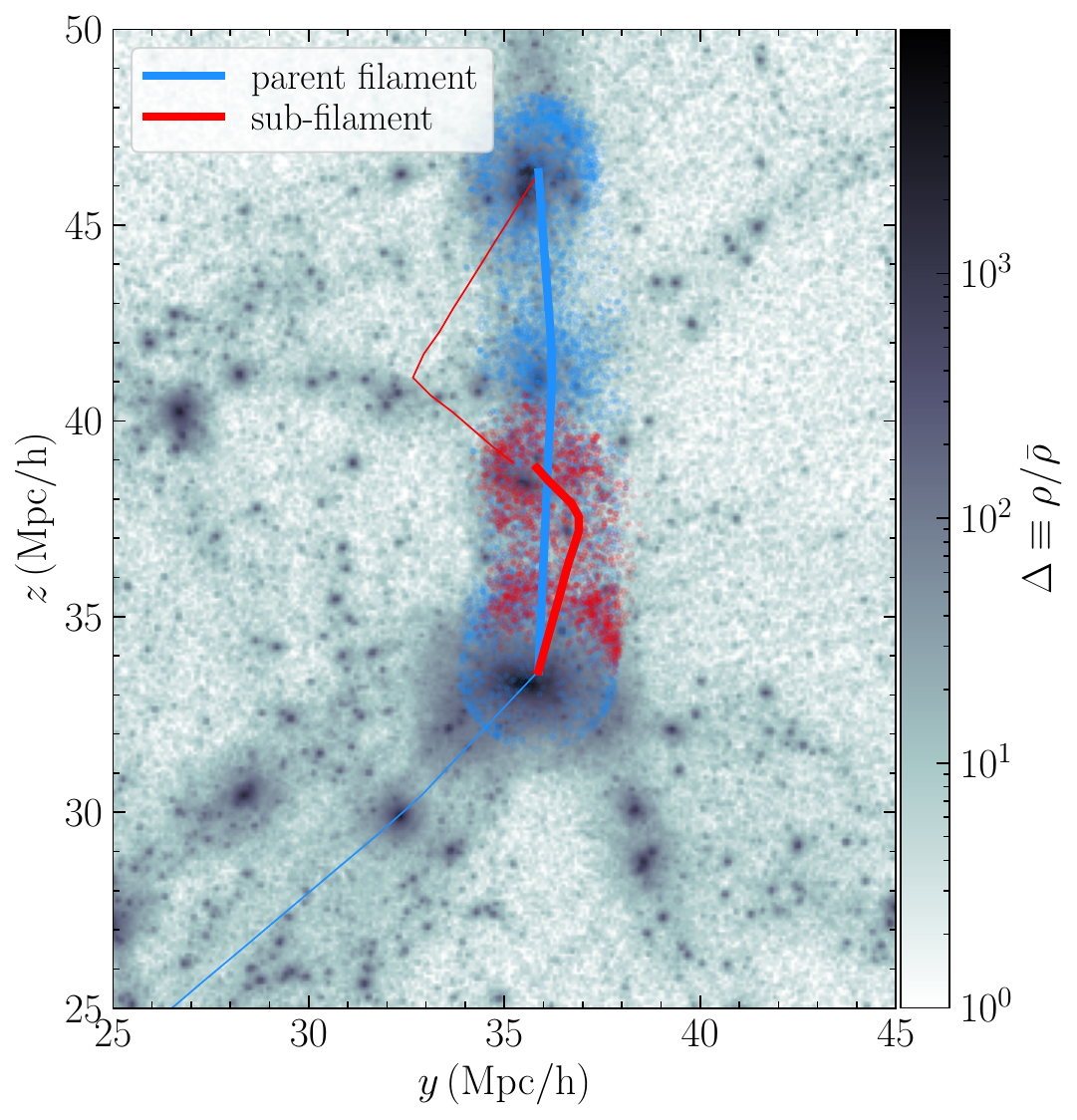}

\caption{Visualization of parent filaments (blue) and sub-filaments (red) in a $10\Mpch$-thick slice. We focus on one parent/sub-filament pair (thick lines) in the dark matter density slice. The two share a node of mass $4.7 \times 10^{14}\Mh$. The parent connects this node to another of mass $1.4 \times 10^{14}\Mh$, whereas the sub-filament connects it to a node of mass $3.4 \times 10^{13}\Mh$. Their lengths are $12.8$ and $5.8\Mpch$, respectively, while their radii $R_v$ are $4.1$ and $3.8\Mpch$. A fraction of randomly chosen particles within $R_v/3$ of each spine are highlighted. To visually focus on the filaments, we exclude particles lying within the radii of halos resolved with $\geq30$ particles. The second sub-filament (thin red line) is also associated with the same main filament. The example highlights the idea of sub-filaments as distinct objects residing in the (stronger) tidal environment of their parent filaments.
}
\label{fig:visual}
\end{figure}

Some of these candidates are not physically distinct structures but simply repeated detections of the same filament arising from the use of different tracer populations. To identify such cases, we perform a second comparison using a much smaller matching radius corresponding to the uncertainty in the reconstructed spine position.
This uncertainty is estimated from the mean displacement, $d_{\rm s}$, between the Fourier-smoothed filament spine and the final spine obtained after the transverse mass asymmetry minimization. 
If more than $80\%$ of a candidate sub-filament spine lies within $3\times d_{\rm s}$ of a previously identified filament, the two are taken to represent the same physical structure, and the shorter resampled filament is removed from the filament catalogue.
The remaining candidates are classified as genuine \emph{sub-filaments} while the remaining filaments -- that are not embedded within larger systems -- are denoted \emph{parent} filaments. 

A representative parent and sub-filament pair identified by \skeletor\ is shown in Fig~\ref{fig:visual}. The embedded geometry illustrates the different environments in which the two structures reside.

\vskip 0.1in \noindent
\textbf{\emph{IV. \underline{Sub-filament properties:}}}\\
Having identified the sub-filament population, we now compare some of its basic properties with those of the parent filaments. The goal is to determine whether subfilaments represent a statistically distinct subset or simply a random sample of the filament population.

\begin{figure}
\centering
\includegraphics[width=\linewidth,trim=6 10 6 6,clip]{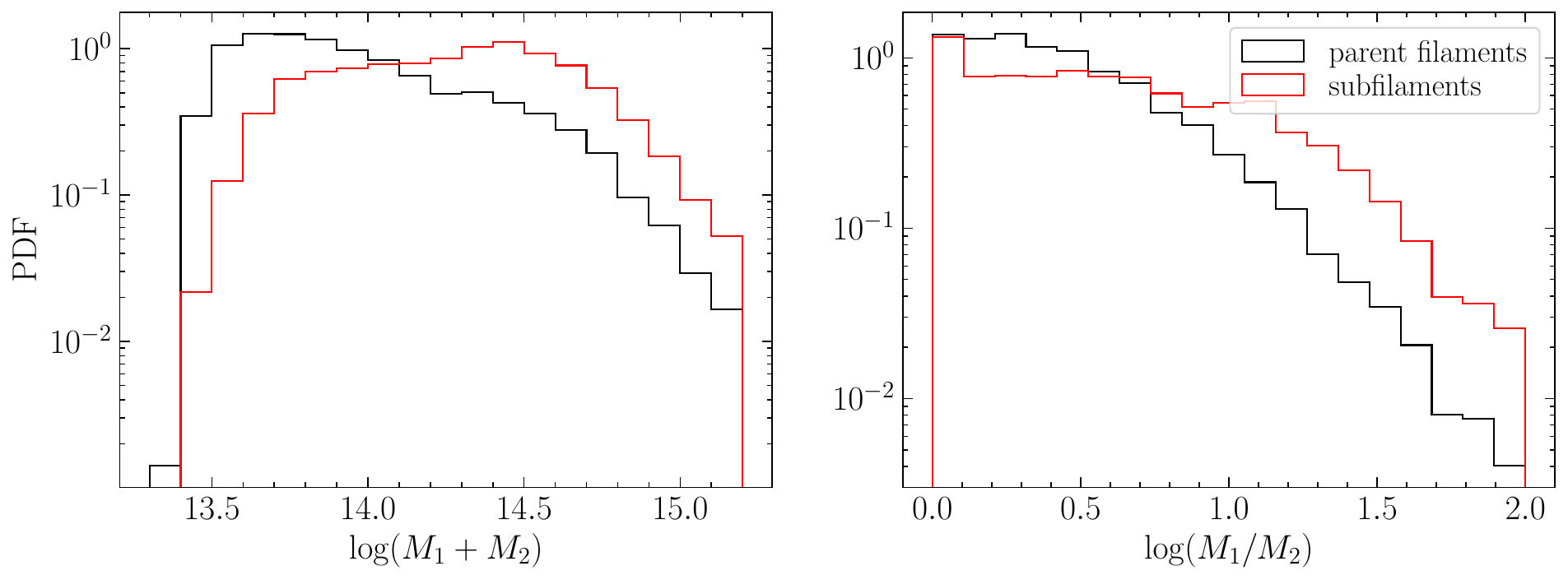}
\caption{\emph{Left panel:} Distribution of the total node mass, $\log(M_1+M_2)$. \emph{Right panel:} Distribution of the corresponding node mass ratio, $M_1/M_2$. Parent filaments are shown in black and sub-filaments in red. Node masses are reported in \Mh.}
\label{fig:node_mass_parentsub}
\end{figure}

The \emph{left panel} of Fig.~\ref{fig:node_mass_parentsub} shows the distribution of the total node mass, $\log(M_1+M_2)$. Subfilaments are preferentially associated with larger total node masses, implying that they are more commonly found in the vicinity of massive nodes and in denser environments. This behaviour is expected if hierarchical substructure becomes increasingly abundant around the most prominent filaments of the cosmic web. The higher, rarer node masses also suggest that sub-filaments often share a node with their parent filaments. 

The \emph{right panel} of Fig.~\ref{fig:node_mass_parentsub} compares the distribution of the node mass ratio, $M_1/M_2$, for parent filaments and sub-filaments, where $M_1 \geq M_2$ are the masses of the two nodes connected by a filament. The distribution exhibits a more prominent high-ratio tail for the sub-filament population, indicating that sub-filaments preferentially connect nodes with more unequal masses. This supports the node-sharing hypothesis mentioned above, where sub-filaments are sub-structures of major filaments, connecting to a massive node at one end belonging to the parent filament, while terminating within the body of the parent filament rather than at another comparably massive node. A detailed examination of individual parent/sub-filament pairs to test this interpretation is left for future work.

\begin{figure}
\centering
\includegraphics[width=\linewidth, trim=13 10 10 4,clip]{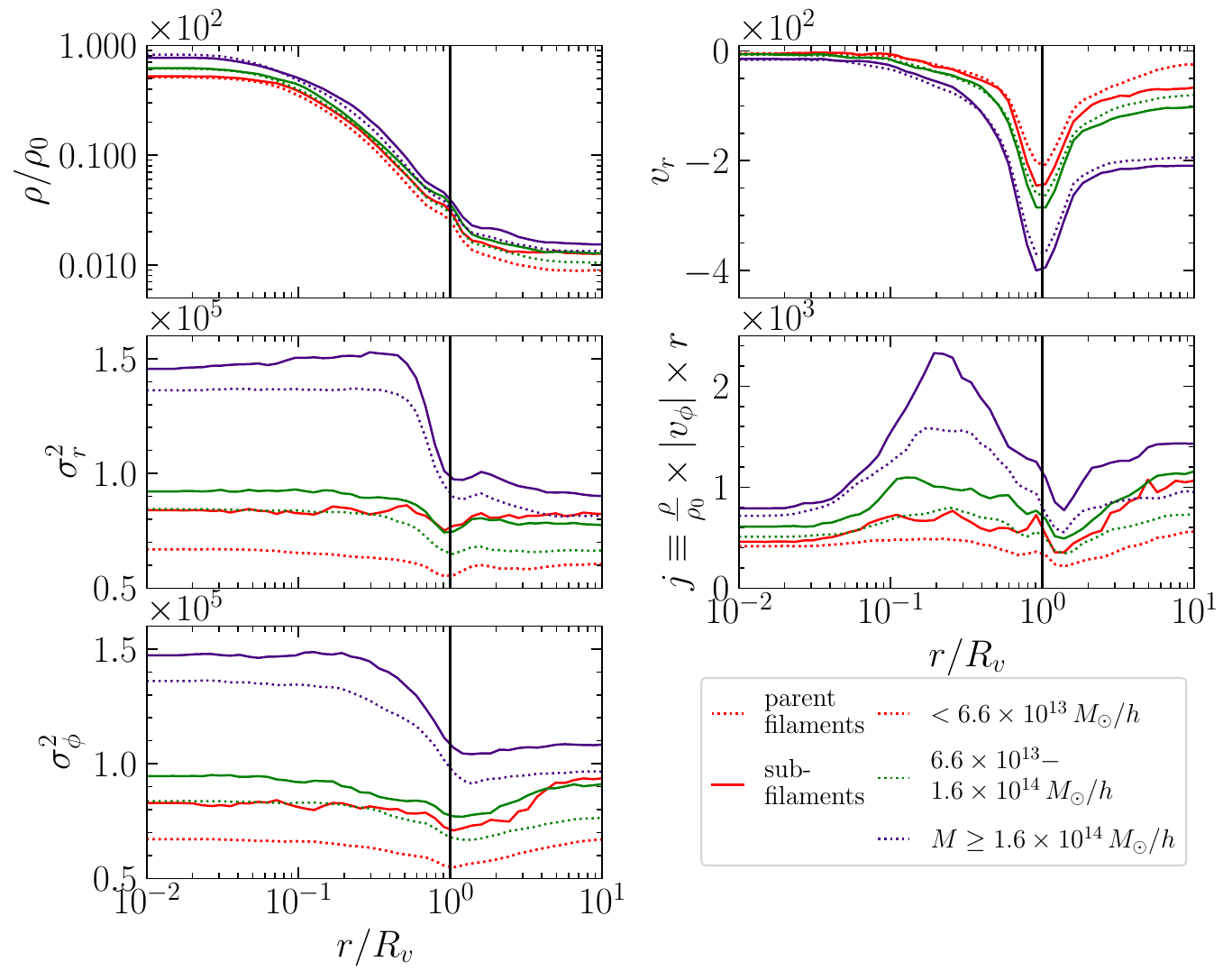}
\caption{Comparison of the radial density and velocity profiles of parent filaments (dotted curves) and subfilaments (solid curves). Different colours correspond to different bins in the total node mass, defined by the $33.33$ and $66.67$ percentiles of the full filament population. The radial distance is scaled by the radius of the filament,  $R_v$. $v_r$ ($j$) is given in units of km/s (km \Mpch/s), and the velocity dispersions are reported in $(\mathrm{km/s})^2$. Vertical black line marks $r=R_v$.}
\label{fig:parent_sub_profs_mnode_split}
\end{figure}

To study whether these environmental differences are reflected in the internal properties of the filaments, we compare radial profiles of the density and velocity fields for parent filaments and sub-filaments. We focus on five quantities: the density normalized by the mean background density $\rho/\rho_0$, the radial infall velocity $v_r$, where $r$ is the perpendicular distance to the spine,
the absolute axial angular momentum $j \equiv \rho/\rho_0 \times |v_\phi| \times r$, and the dispersion of radial and tangential velocities: $\sigma^2_r, \sigma^2_\phi$. When constructing the stacked profiles, each filament is weighted by the length over which its profile is measured. This effectively yields volume-weighted averages, as the area of cross-section is the same for all the filaments in a given radial bin.

Since many filament properties depend strongly on node mass (cf., paper II), the comparison is performed within fixed bins of total node mass. The radial distance is scaled by the radius $R_v$ of each filament inferred from the velocity dip, which accounts for most of the radial dependence. The filaments are divided into three bins sorted by total node mass and containing approximately equal numbers of filaments. The resulting profiles are shown in Fig.~\ref{fig:parent_sub_profs_mnode_split}.

At the lowest node masses, the density profiles reveal significantly denser outskirts  ($r/R_v>1$) of sub-filaments, by a factor $\sim1.5$, as compared to parent filaments of the same node mass (red curves in the \emph{upper left panel}). This difference substantially weakens at higher node masses, although it is still visible (purple and green curves in the same panel).

The radial velocity profiles (\emph{upper right panel}) exhibit a similar trend. Sub-filaments show a deeper infall dip than parent filaments, although the location of the minimum remains nearly unchanged. The differences are again most evident for the lowest node masses in the outskirts, where stronger infall persists out to larger radii for the sub-filament population. This suggests that sub-filaments continue to accrete matter in anisotropic environments influenced by nearby overdensities and the gravitational potential of larger structures.

The axial angular momentum profiles (\emph{middle right panel}) also differ systematically between the two populations. At fixed (especially, low) node mass, sub-filaments exhibit $\gtrsim2\times$ larger angular momentum 
than parent filaments, implying stronger rotations in their surroundings. Finally, both the radial and tangential velocity dispersions (\emph{middle} and \emph{lower left panels}, respectively) are consistently higher for sub-filaments, now across all node mass bins, indicating a dynamically hotter environment.

Overall, these comparisons show that sub-filaments differ from parent filaments in several important respects. They preferentially connect more asymmetric and more massive node pairs and exhibit systematically enhanced densities, stronger coherent inflows, larger tangential motions, and higher velocity dispersions even after controlling for node mass and filament radius. We discuss the implications of these trends below.

\vskip 0.1in \noindent
\textbf{\emph{V. \underline{Discussion and future prospects:}}}\\
The results presented here, made possible by our novel use of \skeletor, suggest that parent and sub-filaments represent physically distinct populations. Even after controlling for node mass, sub-filaments exhibit systematically higher densities, stronger radial infall, larger tangential velocities, and higher velocity dispersions. This naturally raises the possibility that treating the two populations separately may provide a more physically meaningful statistical description of the cosmic web.

As mentioned earlier, one possible application of the parent/sub-filament classification is in the search for universality in filament (phase-space) profiles. 
Establishing whether the filamentary hierarchy is 
universal is important for reasons beyond simply obtaining convenient fitting functions \citep{Espinosa+2022}.
Universality, if it exists, would provide valuable clues about the physical processes that govern the formation and evolution of cosmic filaments and the extent to which these processes are insensitive to the details of the local environment. It would also provide a well-defined baseline against which the effects of different cosmological models, dark matter scenarios, or baryonic physics could be compared in a more robust and systematic manner.
The hierarchical nature of the filament population may itself be one source of the large scatter reported in previous studies. If parent and sub-filaments evolve under different physical conditions, combining them into a single sample may therefore wash out otherwise simple scaling relations. A natural next step is therefore to search for universal profiles within each population separately.

The parent/sub-filament classification also raises a number of questions about how (sub-)filaments form and evolve \citep[see, e.g.,][]{jhee+26}. Embedded within larger filamentary systems, sub-filaments will experience strong directional tidal fields and the coherent mass transport taking place along the parent filament. It is therefore natural to expect their evolution to differ from that of isolated filaments. For example, it is not clear whether sub-filaments form independently and are later incorporated into larger structures, or whether they emerge directly as secondary branches during the assembly of the parent filament itself. The subsequent evolution of these structures is equally uncertain. Strong inflows along the parent filament may continuously feed sub-filaments, but they may also disrupt them through tidal interactions.
Establishing the lifetimes of sub-filaments as compared to their parents remains an open question.

The presence of sub-filaments may also modify the way matter is transported through the filament network. Rather than flowing directly along a single spine towards a massive node, material may preferentially flow along multiple smaller sub-filaments. 
Filamentary sub-structure might exhibit systematically different multi-stream regions, phase-space caustics, or filament boundaries compared to the parent population.
Addressing these questions will require tracking individual sub-filaments across cosmic time, an exercise we will undertake in the near future. 

The existence of a sub-filament population may also have implications for the connection between galaxies and the cosmic web. In the halo paradigm, distinguishing between central and satellite galaxies has proven essential for accurately describing galaxy clustering and the galaxy-halo connection, reflecting the fact that galaxies residing in different environments evolve differently \citep{Zehavi+2005, Zehavi+2011}. A similar situation may arise for filaments. Galaxies embedded within sub-filaments are expected to experience a different large-scale environment from those associated with isolated parent filaments, including stronger tidal fields and the coherent mass transport taking place within the surrounding host filament, possibly leading to imprints in gas accretion and star formation history, merger rates or spatial alignments.
If such trends exist, explicitly accounting for the filament hierarchy may augment existing models based primarily on halo properties.

\vskip 0.1in\noindent
Our analysis has revealed, for the first time, an enigmatic population of \emph{sub-filaments} in the cosmic web. Studying their abundance, spatial distribution, lifetimes and evolution can not only help establish their role in structure formation, but also aid in the search for a universal cosmic hierarchy with implications for cosmology, dark matter phenomenology and galaxy formation.
We hope this motivates the community to explore hierarchical classifications more systematically in future studies of the cosmic web.

\vskip 0.1in \noindent
\textbf{\emph{ Acknowledgments:}} We gratefully acknowledge the use of high performance computing facilities at IUCAA, Pune. AP thanks Ravi Sheth for useful discussions.

\bibliography{references}

\end{document}